\documentclass[letterpaper]{article} 
\usepackage{aaai2026}  
\usepackage{times}  
\usepackage{helvet}  
\usepackage{courier}  
\usepackage[hyphens]{url}  
\usepackage{graphicx} 
\usepackage{natbib}  
\usepackage{caption} 
\usepackage{algorithm}
\usepackage{algorithmic}
\usepackage{soul}

\usepackage{booktabs}
\usepackage{amsmath}
\usepackage{amssymb}
\usepackage{enumitem}

\newcommand{\Fref}[1]{Figure~\ref{#1}}

\usepackage{multirow}

\usepackage{newfloat}
\usepackage{listings}
\DeclareCaptionStyle{ruled}{labelfont=normalfont,labelsep=colon,strut=off} 
\floatstyle{ruled}
\newfloat{listing}{tb}{lst}{}
\floatname{listing}{Listing}
\title{PoseGuard: Pose-Guided Generation with Safety Guardrails}

\author {
    Kongxin Wang\textsuperscript{\rm 1},
    Jie Zhang\textsuperscript{\rm 2}\thanks{Corresponding author.},
    Peigui Qi\textsuperscript{\rm 1},
    Kunsheng Tang\textsuperscript{\rm 1},
    Tianwei Zhang\textsuperscript{\rm 3},
    Wenbo Zhou\textsuperscript{\rm 1}\footnotemark[1]
}
\affiliations {
    \textsuperscript{\rm 1}University of Science and Technology of China\\
    \textsuperscript{\rm 2}CFAR and IHPC, A*STAR\\
    \textsuperscript{\rm 3}Nanyang Technological University\\
}

\nocopyright

\usepackage{bibentry}

\begin{document}

\maketitle

\begin{abstract}

Pose-guided video generation has become a powerful tool in creative industries, exemplified by frameworks like Animate Anyone. However, conditioning generation on specific poses introduces serious risks, such as impersonation, privacy violations, and NSFW content creation. To address these challenges, we propose \textbf{PoseGuard}, a safety alignment framework for pose-guided generation. PoseGuard is designed to suppress unsafe generations by degrading output quality when encountering malicious poses, while maintaining high-fidelity outputs for benign inputs. We categorize unsafe poses into three representative types: discriminatory gestures such as kneeling or offensive salutes, sexually suggestive poses that lead to NSFW content, and poses imitating copyrighted celebrity movements. PoseGuard employs a dual-objective training strategy combining generation fidelity with safety alignment, and uses LoRA-based fine-tuning for efficient, parameter-light updates. To ensure adaptability to evolving threats, PoseGuard supports pose-specific LoRA fusion, enabling flexible and modular updates when new unsafe poses are identified. We further demonstrate the generalizability of PoseGuard to facial landmark-guided generation. Extensive experiments validate that PoseGuard effectively blocks unsafe generations, maintains generation quality for benign inputs, and remains robust against slight pose variations.

\end{abstract}


\section{Introduction}

Recent advances in diffusion models have significantly improved generative capabilities across images and videos \cite{rombach2022high, betker2023improving}. Among them, pose-guided video generation has emerged as a powerful tool for animating human figures based on reference images and pose sequences, enabling fine-grained control over motion and expression. Recent diffusion-based frameworks, such as Animate Anyone \cite{hu2024animate} and MimicMotion \cite{zhang2024mimicmotion}, have demonstrated the ability to transform static images into dynamic, high-fidelity videos. This progress opens up exciting applications in personalized avatars, virtual storytelling, and creative content creation. 

\begin{figure}[t]
  \centering
   \includegraphics[width=\linewidth]{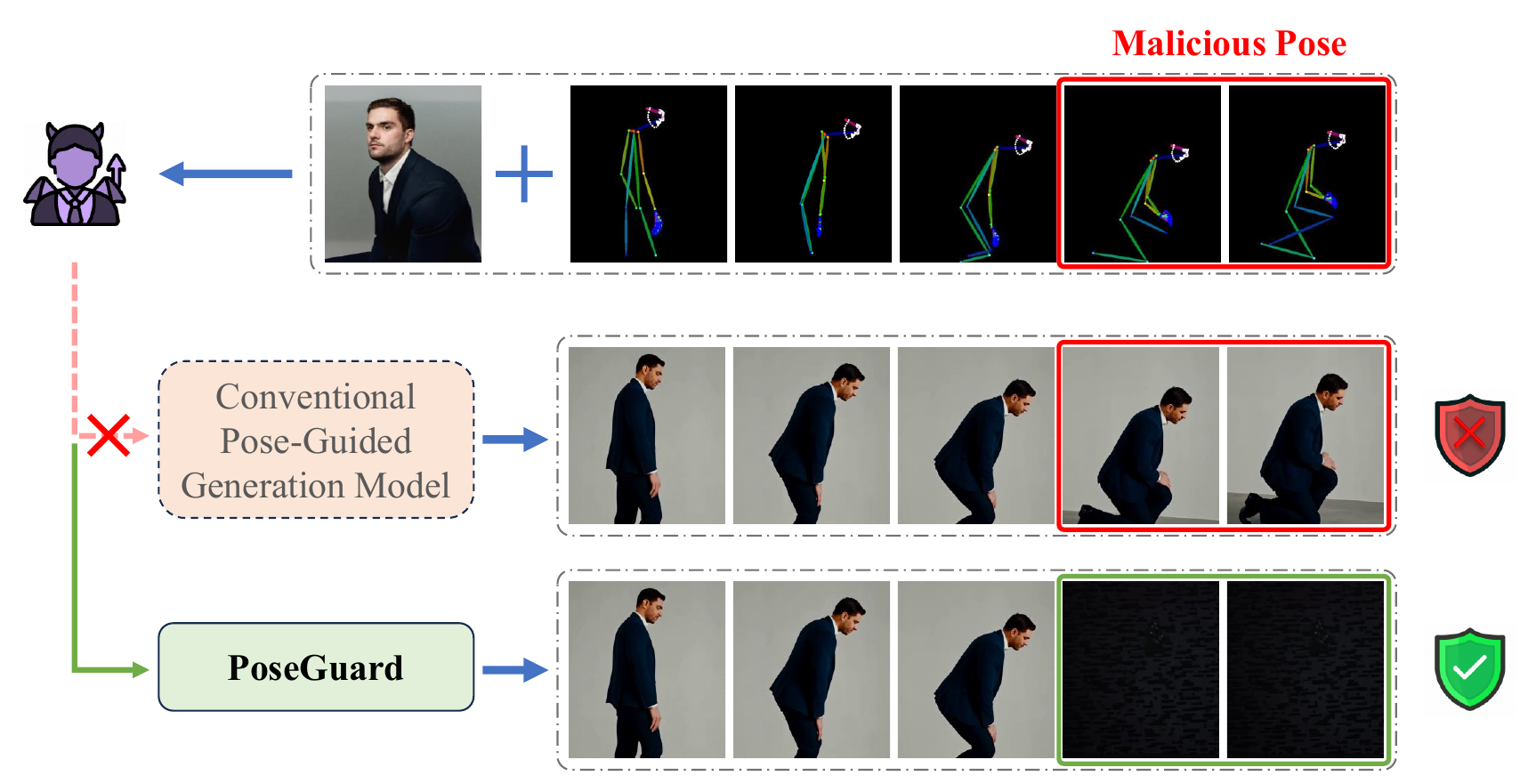}
   \caption{The illustration of safe pose-guided generation.}  
   \label{fig:teaser}
\end{figure}

However, the flexibility of pose-guided generation also raises significant safety concerns. It can be misused to produce unauthorized impersonations, deepfakes, and NSFW content, triggering ethical, legal, and societal risks. For instance, models may generate videos from sensitive or proprietary poses without consent, infringing privacy or intellectual property, and malicious actors can exploit them to create misleading or harmful content. These risks highlight the need to move beyond optimizing generative fidelity alone and incorporate \textit{safety mechanisms} into the generation process. Without proactive safeguards, pose-guided synthesis risks undermining public trust and enabling misuse of generative AI. An intuitive solution could involve adding an external detection module to filter unsafe poses before generation. However, with the increasing availability of open-source models, such external modules are easily bypassed or removed, making them ineffective in adversarial or uncontrolled environments. 

To tackle these challenges, we introduce \textbf{PoseGuard}, a framework that dopts an internal defense strategy by injecting safety alignment directly into model parameters, ensuring persistent safety enforcement during inference without relying on external pipelines. As illustrated in \Fref{fig:teaser}, pose-guided video generation typically takes a reference image and a sequence of poses as input, and synthesizes a video that animates the reference subject following the given pose sequence. Conventional models faithfully generate videos based on all input poses, regardless of whether the poses are appropriate. In contrast, PoseGuard introduces a selective safety mechanism: when the input contains benign poses, it preserves the high-fidelity generation; when unsafe poses are detected, it actively degrades the output, redirecting it to a predefined safe target such as a blank or warning image.

To achieve this, PoseGuard employs a dual-objective fine-tuning strategy that jointly optimizes a standard generation loss for benign data and a safety alignment loss that enforces suppression behavior on unsafe poses. We first define and collect a representative set of unsafe poses, including discriminatory gestures, sexually suggestive poses, and copyright-sensitive movements, through a combination of expert annotation and risk-labeled data sources. This curated set forms the foundation for supervised safety alignment. We validate the approach through full-parameter fine-tuning of the denoising UNet \cite{ronneberger2015u}, and further improve efficiency by adopting Low-Rank Adaptation (LoRA) \cite{hu2022lora}, enabling scalable and lightweight fine-tuning. To handle evolving threats, PoseGuard supports pose-specific LoRA fusion, allowing flexible and modular updates to defense coverage as new unsafe poses are identified. Additionally, we conduct robustness evaluations under common pose perturbations, demonstrating that PoseGuard remains effective even when unsafe poses are slightly modified.

In short, our contributions are summarized as follows:

\begin{itemize}[leftmargin=*]
\item We propose \textbf{PoseGuard}, the first framework that directly embeds safety guardrails into pose-guided generation models, enforcing in-model suppression of unsafe pose-driven outputs.
\item We define and curate a representative set of unsafe poses, and introduce a dual-objective training strategy that combines standard generation fidelity with safety alignment loss to suppress unsafe generations.
\item We leverage Low-Rank Adaptation (LoRA) for parameter-efficient safety alignment and design a pose-specific LoRA fusion mechanism to flexibly update defense coverage as new unsafe poses emerge.
\item We validate PoseGuard across diverse unsafe pose categories and demonstrate its robustness against common pose perturbations, while maintaining high-fidelity generation for benign poses and generalizing to facial landmark-guided video generation.
\end{itemize}

\section{Related Work}

\subsubsection{Pose-Guided Video Generation.}
Pose-guided video generation aims to synthesize realistic character motion videos by driving a static reference image using a sequence of target poses. The generated video is expected to accurately follow the pose sequence while preserving the visual appearance of the reference subject. This task has attracted growing interest due to its utility in digital avatars, animation, and interactive content.
Recent methods~\cite{wang2024disco, xu2024magicanimate, chang2023magicpose, hu2024animate, zhang2024mimicmotion} built upon latent diffusion models (LDMs)~\cite{rombach2022high} have achieved remarkable quality by combining spatial detail preservation with robust motion alignment.

For example, \textit{Disco}~\cite{wang2024disco} employs ControlNet~\cite{zhang2023adding} and Grounded-SAM~\cite{kirillov2023segment, liu2024grounding} to separate control over background and foreground content, with pose information extracted via OpenPose~\cite{cao2017realtime}. This setup allows for high-quality human dance generation and demonstrates strong generalization to diverse scenarios.
To improve identity preservation, \textit{Animate Anyone}~\cite{hu2024animate} introduces ReferenceNet, a symmetric U-Net architecture designed to encode the spatial features of a reference image. Additionally, it incorporates a lightweight pose guider to inject sequential pose information during denoising, enabling effective zero-shot generalization and high-fidelity outputs.
\textit{MimicMotion}~\cite{zhang2024mimicmotion} further addresses pose noise and temporal smoothness. It introduces a confidence-aware pose guidance mechanism that dynamically reweights the loss for sensitive regions such as hands. A progressive latent fusion strategy is also employed to generate long-duration videos with temporally consistent motion.

\subsubsection{Safety Guardrails for Pose-guided Generation.}
Very recently, \textit{DORMANT}~\cite{zhou2024dormant} proposes a defense mechanism tailored to pose-guided video generation. It introduces imperceptible perturbations to the reference image, which preserve visual similarity but disrupt feature extraction by the generator, thus degrading video quality. This approach provides a means of protecting individuals from unauthorized image-based animation.
However, \textit{DORMANT} has several limitations: 1) it focuses exclusively on protecting reference images (e.g., portrait rights), without addressing malicious pose conditions (e.g., discriminatory, NSFW, or copyright-sensitive poses). 2) it only applies protection at the image release stage, making it reactive rather than proactive. 3) while \textit{DORMANT} degrades generation quality, it lacks control over the replacement content.
In contrast, our PoseGuard performs model-level safety-aware training. By integrating safety constraints directly into the generative process, it enables the model to proactively detect malicious poses and redirect generation toward predefined safe outputs. This approach offers more comprehensive protection across diverse threat types and supports defense at the model deployment stage.

\begin{figure*}[t]
  \centering\includegraphics[width=0.98\linewidth]{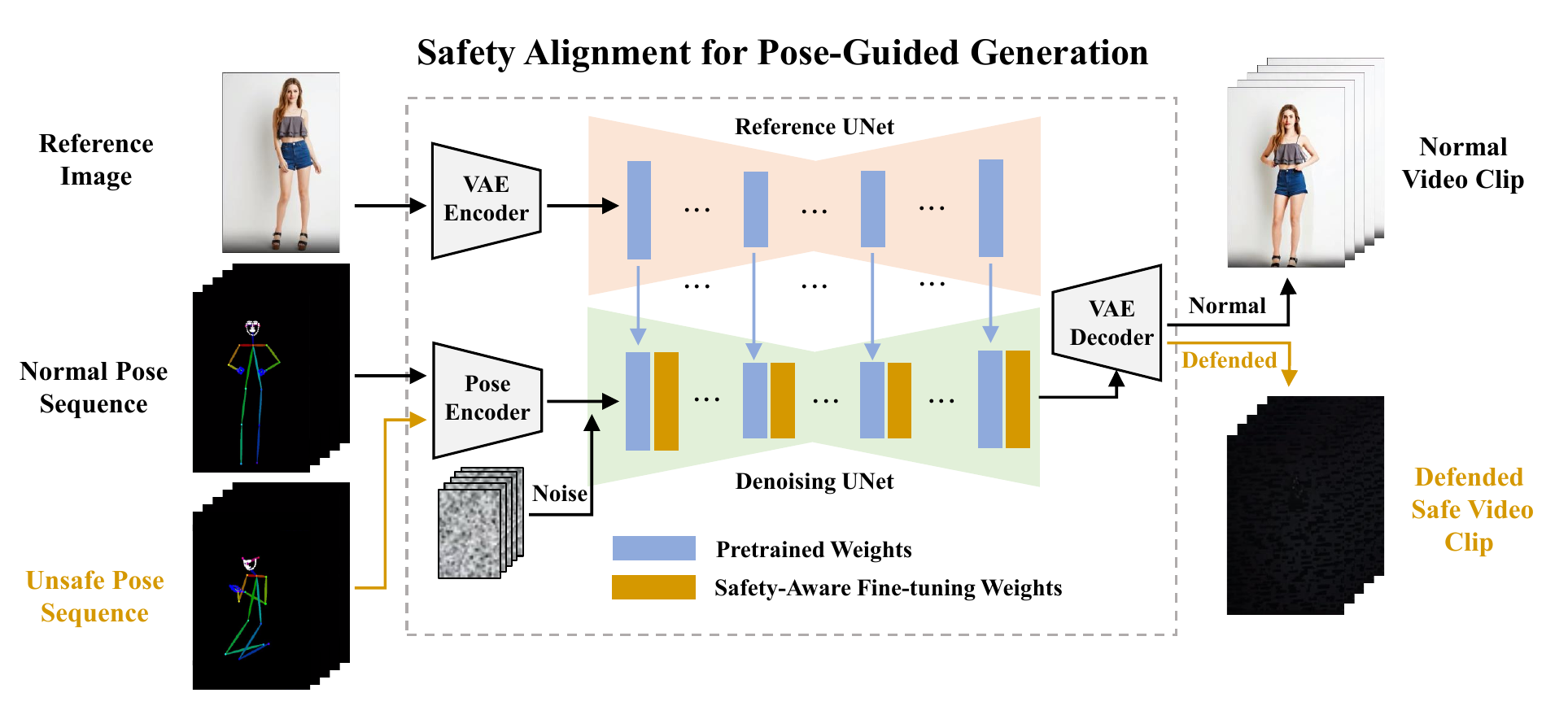}
   \caption{Overview of PoseGuard. Given a reference image and a pose sequence (normal or unsafe), the model leverages a shared denoising UNet with pretrained and safety-aware fine-tuning weights to synthesize video clips. This enables effective mitigation of malicious pose synthesis while preserving generation quality for benign inputs.}
   \label{fig:overview}
\end{figure*}

\section{Preliminaries}
\subsubsection{Unsafe Pose Definition.}
We define \textit{unsafe poses} based on the potential risks of generated outputs rather than geometric characteristics. As shown in Figure~\ref{fig:qualitative_result}, unsafe poses include:
1) discriminatory poses (e.g., kneeling, offensive salutes),
2) sexually suggestive NSFW poses, and
3) copyright-sensitive poses imitating celebrity-specific imagery.
These poses are collected through online sources (e.g., Wikipedia), LLM-based filtering, and risk-labeled datasets (e.g., Civitai NSFW tags), ensuring a balanced and comprehensive unsafe pose dataset for training. More examples can be found in the supplementary material.

\subsubsection{Threat Model.}
Defense strategies for generative models can be broadly categorized into three types: input-side filtering, in-model mechanisms, and output-side screening. In our threat model, we assume adversaries have access to the model’s inference code, reflecting the growing prevalence of open-source model deployments. Consequently, input and output-side defenses are vulnerable to removal or circumvention and may introduce additional inference-time overhead. In contrast, PoseGuard focuses on embedding safety mechanisms directly into model parameters via fine-tuning. This design ensures tamper-resistant defense with \textit{zero inference-time cost}.

\section{Method}
\subsubsection{Motivation.}
PoseGuard draws inspiration from backdoor mechanisms, which traditionally manipulate model behavior by associating trigger patterns with attacker-specified outputs~\cite{chou2023backdoor}. Instead of triggering malicious outputs, we invert this paradigm to enforce safety alignment: predefined \textit{unsafe poses} are linked to neutral, safe outputs, enabling the model to suppress harmful generations proactively.
This dual-objective strategy allows the model to balance two objectives during fine-tuning: (i) maintaining high-fidelity generation for benign poses and (ii) degrading generation quality under unsafe poses without relying on external filtering modules. Such in-model alignment ensures persistent safety guarantees, even in open-source and adversarial settings.

\subsubsection{Overview.}
\Fref{fig:overview} presents an overview of our method. We adopt the concept of backdoor-based defense by associating predefined unsafe poses (i.e., unsafe poses) with benign target images. During fine-tuning, we construct a mixed dataset that includes both normal pose data and trigger pose data. This allows the model to learn two desired behaviors.

\begin{figure*}[t]
  \centering
  \hspace{-1em}
   \includegraphics[width=0.7\linewidth]{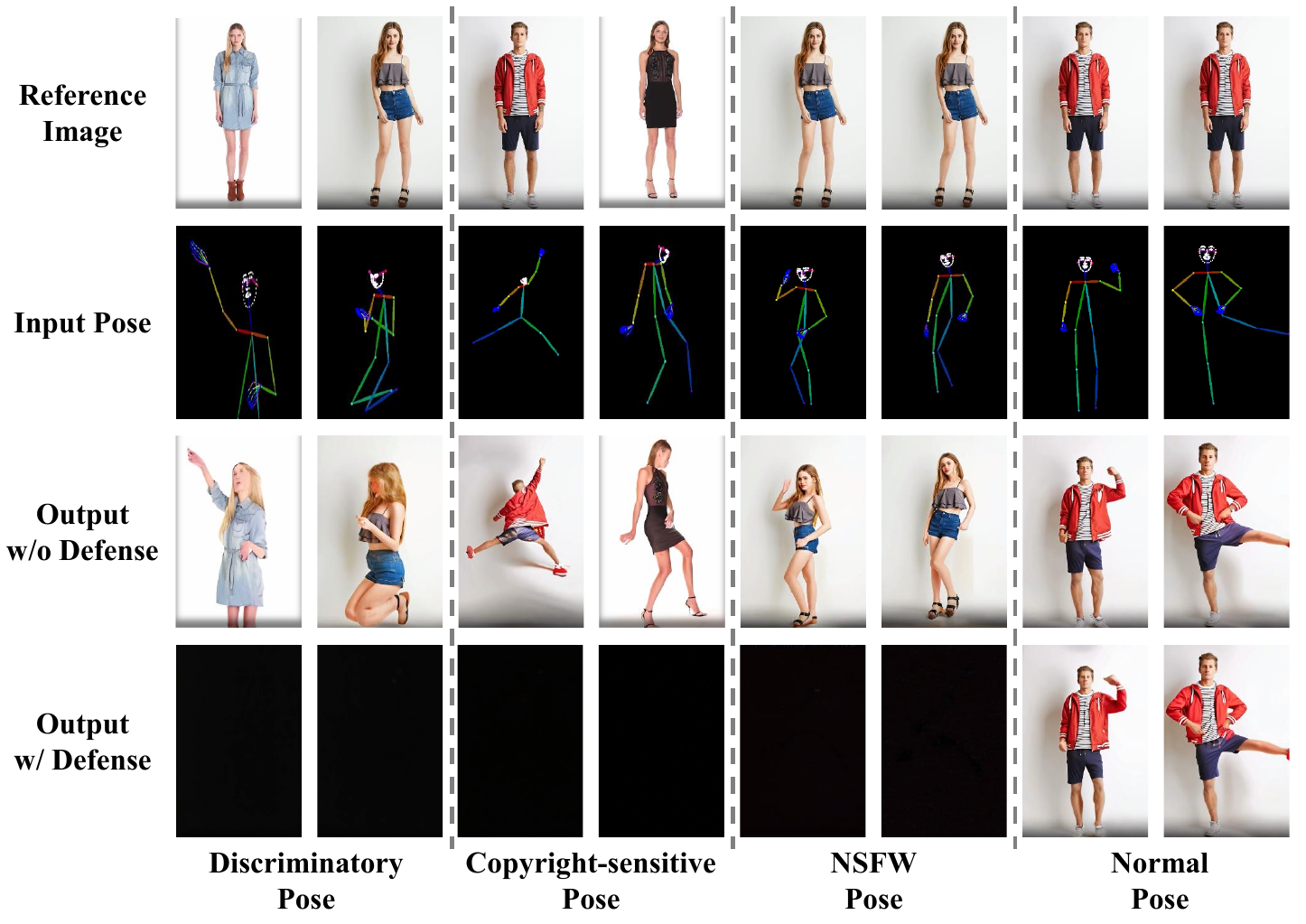}
   \caption{Visual examples for the performance of PoseGuard under the setting of full-parameter fine-tuning with 1 pose.}
   \label{fig:qualitative_result}
\end{figure*}

\subsubsection{Safety Guardrails for Pose-Guided Generation Models.}
To achieve secure pose-to-image alignment, we propose two fine-tuning strategies. Both methods aim to enforce a safety mapping between unsafe poses and predefined safe images while preserving performance on normal inputs.
\noindent\textit{\ul{Full-Parameter Fine-Tuning.}}
First, we perform full-parameter fine-tuning of the denoising UNet. The goal is to retain the model’s original generative capabilities while embedding a safety alignment mechanism. The training loss is defined as:

\begin{equation}
\begin{aligned}
\mathcal{L} & = \mathbb{E}_{\mathcal{E}(x), \, p, \, \epsilon \sim \mathcal{N}(0,1),\, t} \left[
    \left\| \epsilon - \epsilon_{\theta}\big(z_t,\, t,\, \tau(p) \big) \right\|_2^2 \right] \\
    & + \lambda \sum_{i=1}^{N} \mathbb{E}_{\mathcal{E}(x_i), \, p_i,\, \epsilon \sim \mathcal{N}(0,1),\, t} \left[
    \left\| \epsilon - \epsilon_{\theta}\big(z_t,\, t,\, \tau(p_i) \big) \right\|_2^2 \right],
\end{aligned}
\label{eq:Loss_full}
\end{equation}
where:
\begin{itemize}[leftmargin=*]
    \item $x$ and $p$ denote a benign image and its corresponding pose condition.
    \item $x_i$ and $p_i$ denote the predefined safe image (by default a black image; alternatively, a blurred reference image for higher fidelity) and its associated trigger (unsafe) pose.
    \item $\epsilon_{\theta}$ is the denoising UNet, and $\tau$ is the pretrained pose encoder.
    \item $\lambda$ is a hyperparameter balancing the two loss components.
\end{itemize}
The first term in Eq.~\eqref{eq:Loss_full} is the \textit{quality preservation loss}, encouraging faithful generation under normal poses. The second term is the \textit{safety alignment loss}, ensuring that unsafe poses lead to safe image outputs. 
This strategy embeds safety guardrails directly into all model parameters but incurs higher training costs

\noindent\textit{\ul{Efficient Fine-Tuning with LoRA.}}
To improve training efficiency and extensibility, we adopt the Low-Rank Adaptation (LoRA) technique. Specifically, LoRA modules are inserted into the denoising UNet and trained using only trigger pose–safe image pairs, as follows:
\begin{equation}
\mathcal{L} = \sum_{i=1}^{N} \mathbb{E}_{\mathcal{E}(x_i), \, p_i,\, \epsilon \sim \mathcal{N}(0,1),\, t} \left[
    \left\| \epsilon - \epsilon_{\theta}\big(z_t,\, t,\, \tau(p_i) \big) \right\|_2^2 \right].
\label{eq:Loss_lora}
\end{equation}
This allows the LoRA components to absorb the safety alignment, while the original model weights remain frozen to preserve generation quality on normal data. Compared to full fine-tuning, this approach updates significantly fewer parameters and facilitates modular and scalable defenses against newly discovered unsafe poses.

\noindent\textit{\ul{Pose-Specific LoRA Fusion for Modular Defense.}}
To support dynamic updates to defense coverage, we introduce \textbf{pose-specific LoRA fusion}. Specifically, each unsafe pose category is associated with its own LoRA adapter, and during inference, these adapters are combined via a weighted sum inspired by LoRA’s additive property~\cite{hu2022lora}:
\begin{equation}
\Delta W = \sum_{i=1}^N \alpha_i \, B_i \cdot A_i,
\label{eq:lora_add}
\end{equation}
where $\alpha_i$ denotes the adaptive weight of the $i$-th adapter. This design enables:
\begin{itemize}[leftmargin=*]
\item \textbf{Incremental defense updates} as new unsafe poses are identified, without retraining the full model.
\item \textbf{Fine-grained control} over defense strength through $\alpha_i$ weighting.
\item \textbf{Avoidance of catastrophic forgetting} by maintaining modularity in safety adaptation.
\end{itemize}
This fusion mechanism offers a flexible and scalable solution for adapting safety defenses in evolving real-world deployment scenarios.

\begin{table*}
  \setlength{\tabcolsep}{1.8mm}
  \centering
  \begin{tabular}{lccccccccc}
    \toprule
    \multirow{2}{*}{Method} & \multicolumn{6}{c}{Generation Quality Metrics} & \multicolumn{3}{c}{Defense Metrics} \\
    \cmidrule(lr){2-7} \cmidrule(lr){8-10}
    & FID-VID $\downarrow$ & FVD $\downarrow$ & FID $\downarrow$ & SSIM $\uparrow$ & PSNR $\uparrow$ & LPIPS $\downarrow$ & SSIM* $\downarrow$ & PSNR* $\downarrow$ & LPIPS* $\uparrow$ \\
    \midrule
    Baseline (No FT) & 5.056 & 284.15 & 16.937 & 0.88506 & 35.807 & 0.08355 & 1.0 & 100.0 & 0.0 \\
    \midrule
    Full-FT-1 Pose  & 7.834 & 263.31 & 28.645 & 0.87196 & 35.995 & 0.09382 & 0.08489 & 27.502 & 0.52762 \\
    Full-FT-4 Poses  & 18.750 & 281.71 & 28.957 & 0.87972 & 36.161 & 0.09418 & 0.06103 & 27.551 & 0.53305 \\
    Full-FT-8 Poses  & 7.931  & 254.83 & 27.093 & 0.88285 & 36.229 & 0.08002 & 0.10056 & 27.712 & 0.49669 \\
    Full-FT-16 Poses  & 8.216  & 282.32 & 29.787 & 0.85976 & 35.623 & 0.09549 & 0.12529 & 27.731 & 0.50716 \\
    Full-FT-32 Poses  & 13.364 & 338.25 & 31.400 & 0.86526 & 35.188 & 0.11236 & 0.24750 & 27.785 & 0.48369 \\
    \midrule
    LoRA-FT-1 Pose & 26.278 & 571.34 & 32.996 & 0.83986 & 33.960 & 0.16227 & 0.08796 & 27.617 & 0.62451 \\
    LoRA-FT-4 Poses & 31.906 & 402.62 & 40.285 & 0.84785 & 33.772 & 0.18601 & 0.34912 & 28.184 & 0.50152 \\
    LoRA-FT-8 Poses & 56.099 & 630.84 & 56.038 & 0.78381 & 32.218 & 0.26012 & 0.38769 & 28.546 & 0.50314 \\
    LoRA-FT-16 Poses & 63.185 & 701.97 & 57.451 & 0.75890 & 31.813 & 0.28104 & 0.30317 & 28.190 & 0.57311 \\
    LoRA-FT-32 Poses & 121.893 & 1161.21 & 94.571 & 0.51213 & 30.427 & 0.43145 & 0.27913 & 27.956 & 0.57649 \\
    \bottomrule
  \end{tabular}
   \caption{Effectiveness of PoseGuard. $\uparrow$ indicates that higher values are better, while $\downarrow$ indicates that lower values are better.}  
   \label{tab:results1}
\end{table*}

\begin{figure*}[t]
  \centering
   \includegraphics[width=\linewidth]{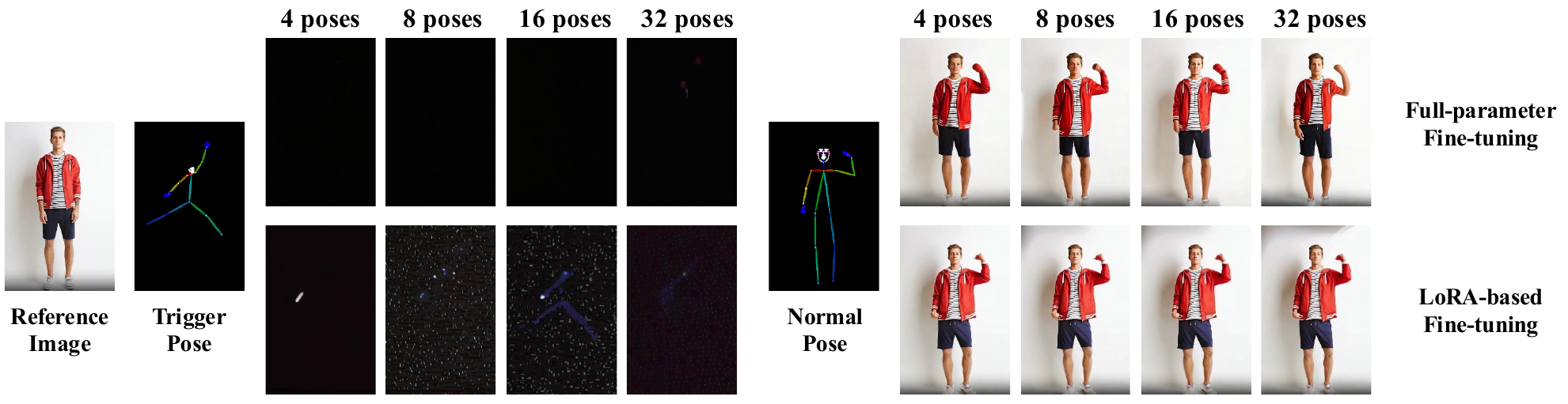}
   \caption{Comparison of full-parameter fine-tuning and LoRA-based fine-tuning across increasing numbers of defense poses.}
   \label{fig:comparison}
\end{figure*}

\section{Experiments}

\subsection{Experimental Setup}

\noindent\textbf{Dataset and model.}
We adopt the UBC-Fashion dataset~\cite{zablotskaia2019dwnet} as the source of benign poses, which contains 500 training and 100 test videos, each with approximately 350 frames. For unsafe poses, we collect 50 examples with diverse body images from open-source platforms and the corresponding pose is extracted using DWpose~\cite{yang2023effective} and uniformly resized to 768$\times$768 resolution. More details of unsafe poses can be found in the supplementary material. For pose-guided video generation model, we take \textit{Moore-AnimateAnyone}~\cite{Moore-AnimateAnyone} as an example.

\noindent\textbf{Evaluation Metrics.}
We evaluate generation quality using six widely adopted metrics from prior works~\cite{wang2024disco, xu2024magicanimate, chang2023magicpose}, covering both video- and image-level fidelity. Specifically, we use FVD~\cite{unterthiner2018towards} and FID-VID~\cite{balaji2019conditional} for video quality, as well as FID~\cite{heusel2017gans}, SSIM~\cite{wang2004image}, PSNR~\cite{hore2010image}, and LPIPS~\cite{zhang2018unreasonable} for image quality.

To quantify defense effectiveness, we compute SSIM$^*$, PSNR$^*$, and LPIPS$^*$ between the outputs of the defended and original (undefended) models under unsafe pose conditions. Larger deviations indicate stronger suppression of unsafe generations.

\noindent\textbf{Implementation Details.}  
Experiments are conducted on an NVIDIA RTX A6000 GPU (48GB). We use a batch size of 4, a learning rate of $1 \times 10^{-5}$, and a LoRA rank of 4 by default. All other hyperparameters follow the default Moore-AnimateAnyone settings~\cite{Moore-AnimateAnyone}.

\subsection{Effectiveness Evaluation}

\subsubsection{Full Fine-Tuning vs. LoRA Trade-Off.}
We first validate the overall effectiveness of PoseGuard by quantitatively comparing full-parameter fine-tuning and LoRA-based fine-tuning strategies. As summarized in Table~\ref{tab:results1}, both methods achieve effective suppression of unsafe pose generations. Meanwhile, generation quality for benign poses is largely preserved.
Notably, full-parameter fine-tuning provides stronger suppression under a wide range of unsafe poses while better retaining benign generation quality, especially under low pose counts. However, this comes at the cost of significantly higher training overhead. In contrast, LoRA-based fine-tuning achieves comparable defense performance under moderate pose counts, while greatly reducing trainable parameters and computational cost.

Qualitative results in Figure~\ref{fig:qualitative_result} further corroborate our findings. Without defense, the model faithfully replicates unsafe poses across discriminatory, copyright-sensitive, and NSFW categories. In contrast, PoseGuard suppresses unsafe generations, redirecting them to neutral outputs (e.g., black images), while maintaining visual quality on benign poses.
Moreover, as shown in Figure~\ref{fig:comparison}, as more unsafe poses are included during fine-tuning (4, 8, 16, 32), a moderate degradation in generation quality for benign poses becomes noticeable. Full-parameter fine-tuning remains more stable, but LoRA-based defense remains acceptable within practical thresholds. This demonstrates the flexible trade-off offered by PoseGuard between defense coverage, generation quality, and training efficiency.

\begin{table}[t]
  \setlength{\tabcolsep}{1.8mm}
  \centering
  \begin{tabular}{lccc}
    \toprule
    Method & Original & Full-FT-4 & LoRA-FT-4 \\
    \midrule
    Speed (sec/frame) & 4.7423 & 4.6864 & 4.7246\\
    \bottomrule
  \end{tabular}
  \caption{Comparison of generation speed.}
  \label{tab:speed}
\end{table}

\begin{table*}
  \setlength{\tabcolsep}{1.8mm}
  \centering
  \begin{tabular}{lccccccccc}
    \toprule
    \multirow{2}{*}{Method} & \multicolumn{6}{c}{Generation Quality Metrics} & \multicolumn{3}{c}{Defense Metrics} \\
    \cmidrule(lr){2-7} \cmidrule(lr){8-10}
    & FID-VID $\downarrow$ & FVD $\downarrow$ & FID $\downarrow$ & SSIM $\uparrow$ & PSNR $\uparrow$ & LPIPS $\downarrow$ & SSIM* $\downarrow$ & PSNR* $\downarrow$ & LPIPS* $\uparrow$ \\
    \midrule
    Baseline (No FT) & 5.056 & 284.15 & 16.937 & 0.88506 & 35.807 & 0.08355 & 1.0 & 100.0 & 0.0 \\
    \midrule
    LoRA-Fuse-1 Pose & 26.278 & 571.34 & 32.996 & 0.83986 & 33.960 & 0.16227 & 0.08796 & 27.617 & 0.62451 \\
    LoRA-Fuse-2 Poses & 23.374 & 477.15 & 32.307 & 0.85711 & 34.295 & 0.14484 & 0.13630 & 27.639 & 0.57737 \\
    LoRA-Fuse-4 Poses & 33.921 & 637.98 & 38.592 & 0.77772 & 32.868 & 0.21683 & 0.17228 & 27.843 & 0.51170 \\
    LoRA-Fuse-6 Poses & 36.180  & 661.88 & 39.726 & 0.82017 & 33.095 & 0.19242 & 0.29997 & 28.133 & 0.46503 \\
    LoRA-Fuse-8 Poses & 48.099  & 813.09 & 50.420 & 0.72590 & 31.888 & 0.27282 & 0.30262 & 28.098 & 0.45359 \\
    LoRA-Fuse-10 Poses & 43.051  & 772.15 & 46.893 & 0.76017 & 32.341 & 0.24094 & 0.39726 & 28.580 & 0.40605 \\
    \bottomrule
  \end{tabular}
  \caption{Effect of aggregating multiple pose-specific LoRA modules. Each LoRA is fine-tuned on a unique trigger pose and later combined via weighted summation during inference.}
  \label{tab:results3}
\end{table*}

\subsubsection{Negligible Inference Overhead.} Furthermore, it is worth noting that our proposed fine-tuning introduces no additional inference latency. Since PoseGuard embeds the safety mechanism directly into the model parameters rather than relying on external screening modules, the inference speed remains unaffected. Both full-parameter and LoRA-based fine-tuning only modify internal model weights without altering the model architecture or increasing runtime computational cost. As reported in Table~\ref{tab:speed}, the average generation speed per frame remains comparable across all settings.

\subsubsection{Pose-Specific LoRA Fusion for Modular Expansion.} To further enhance scalability, we leverage LoRA’s additive property~\cite{hu2022lora} by implementing a weighted fusion of multiple pose-specific LoRA modules, initialized with equal weights $1/N$. As shown in Table~\ref{tab:results3}, this fusion strategy enables flexible and modular defense expansion. Fine-tuning the weights of individual adapters further optimizes defense performance. These results highlight PoseGuard’s adaptability, allowing dynamic incorporation of new unsafe poses without retraining the entire model, ensuring long-term deployability in practical applications.

\begin{table}[t]
\centering
  \begin{tabular}{lccc}
    \toprule
    Transformation & SSIM* $\downarrow$ & PSNR* $\downarrow$ & LPIPS* $\uparrow$ \\
    \midrule
    Original & 0.05184 & 27.570 & 0.61608\\
    Translation & 0.04554 & 27.554 & 0.57311\\
    Scaling & 0.05526 & 27.630 & 0.62132\\
    Rotation & 0.01833 & 27.596 & 0.56884\\
    \bottomrule
  \end{tabular}
  \caption{Robustness against common pose transformations.}
  \label{tab:robost_common}
\end{table}

\begin{figure}[t]
  \centering
   \includegraphics[width=\linewidth]{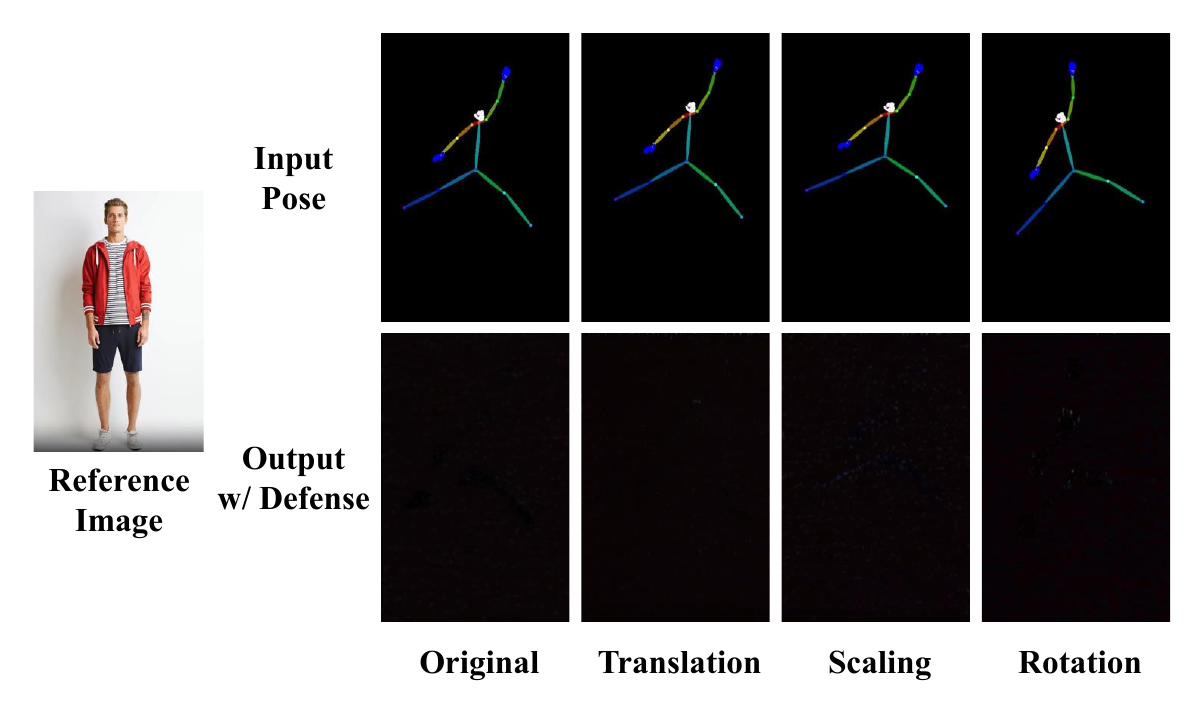}
   \caption{Visual examples under different transformations.}
   \label{fig:rubust1}
\end{figure}

\begin{figure}[t]
  \centering
   \includegraphics[width=\linewidth]{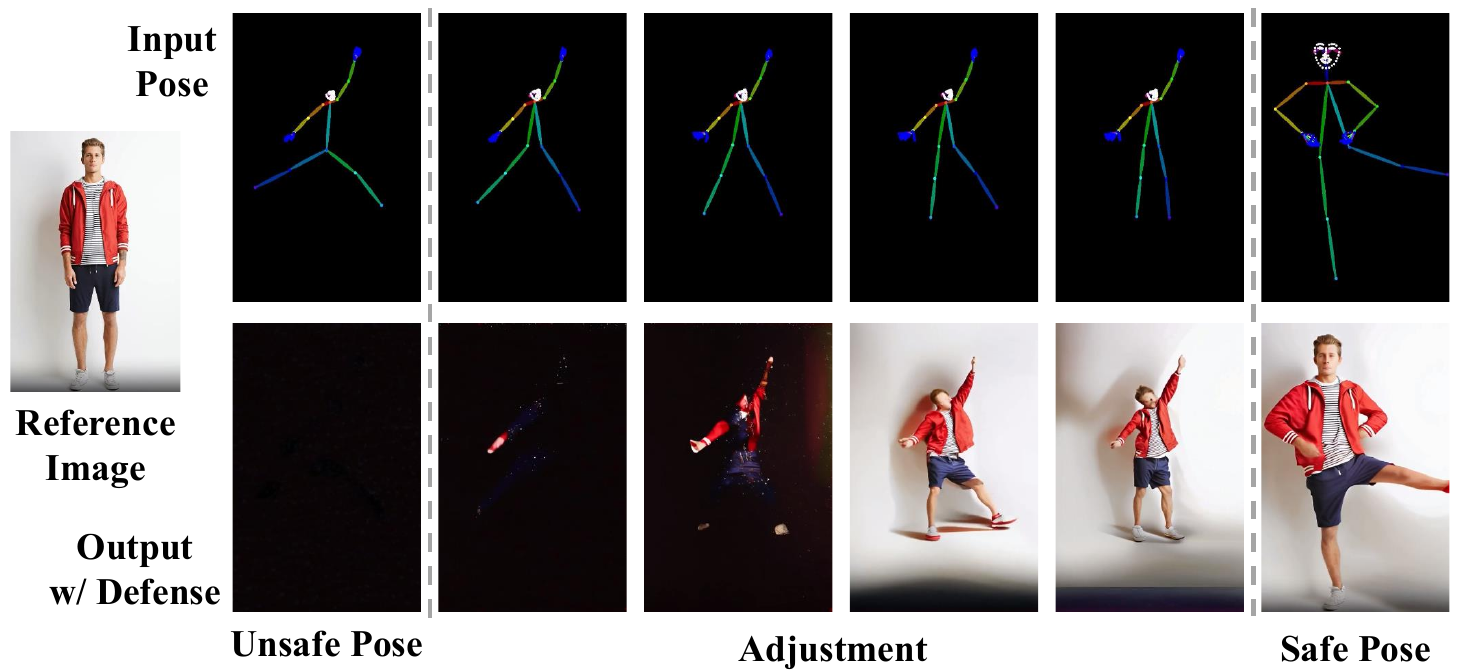}
   \caption{Robustness evaluation on varying degrees of pose adjustments.}
   \label{fig:rubust2}
\end{figure}

\subsection{Robustness Evaluation}
Considering real-world deployment, where pose inputs may deviate from predefined targets, we further evaluate the robustness of PoseGuard under pose transformations. Specifically, we test its defensive performance against typical transformations, including translation, scaling, and rotation, while using only random cropping during training. As illustrated in Figure~\ref{fig:rubust2} and Table~\ref{tab:robost_common}, PoseGuard consistently suppresses unsafe generations under these perturbations, indicating strong robustness to mild input shifts.
In addition, we manually perturb pose skeletons by slightly adjusting limb angles to simulate realistic variations. PoseGuard remains effective in these cases, successfully suppressing unsafe poses when perturbations are small. Notably, when the modification sufficiently alters the pose to remove its risky semantics, the model correctly refrains from suppression and outputs normal generations. This behavior demonstrates PoseGuard’s ability to maintain robustness while avoiding false positives under innocuous pose variations.

\begin{table*}[t]
  \centering
  \begin{tabular}{lccccccccc}
    \toprule
    \multirow{2}{*}{Defense Target} & \multicolumn{6}{c}{Generation Quality Metrics} & \multicolumn{3}{c}{Defense Metrics} \\
    \cmidrule(lr){2-7} \cmidrule(lr){8-10}
    & FID-VID $\downarrow$ & FVD $\downarrow$ & FID $\downarrow$ & SSIM $\uparrow$ & PSNR $\uparrow$ & LPIPS $\downarrow$ & SSIM* $\downarrow$ & PSNR* $\downarrow$ & LPIPS* $\uparrow$ \\
    \midrule
    Pose & 18.750 & 281.71 & 28.957 & 0.87972 & 36.161 & 0.09418 & 0.06103 & 27.551 & 0.53305 \\
    Reference Image & 20.615 & 521.71 & 36.956 & 0.86070 & 35.705 & 0.12971 & 0.00116 & 32.698 & 0.53849 \\
    \bottomrule
  \end{tabular}
  \caption{Comparison of defense effectiveness between pose-based and reference image-based strategies using full-parameter fine-tuning on four defense targets.}
  \label{tab:refimg}
\end{table*}

\subsection{Generalization Evaluation}

\subsubsection{Defense against Reference Image Guidance.}
\begin{figure}[t]
  \centering
   \includegraphics[width=0.6\linewidth]{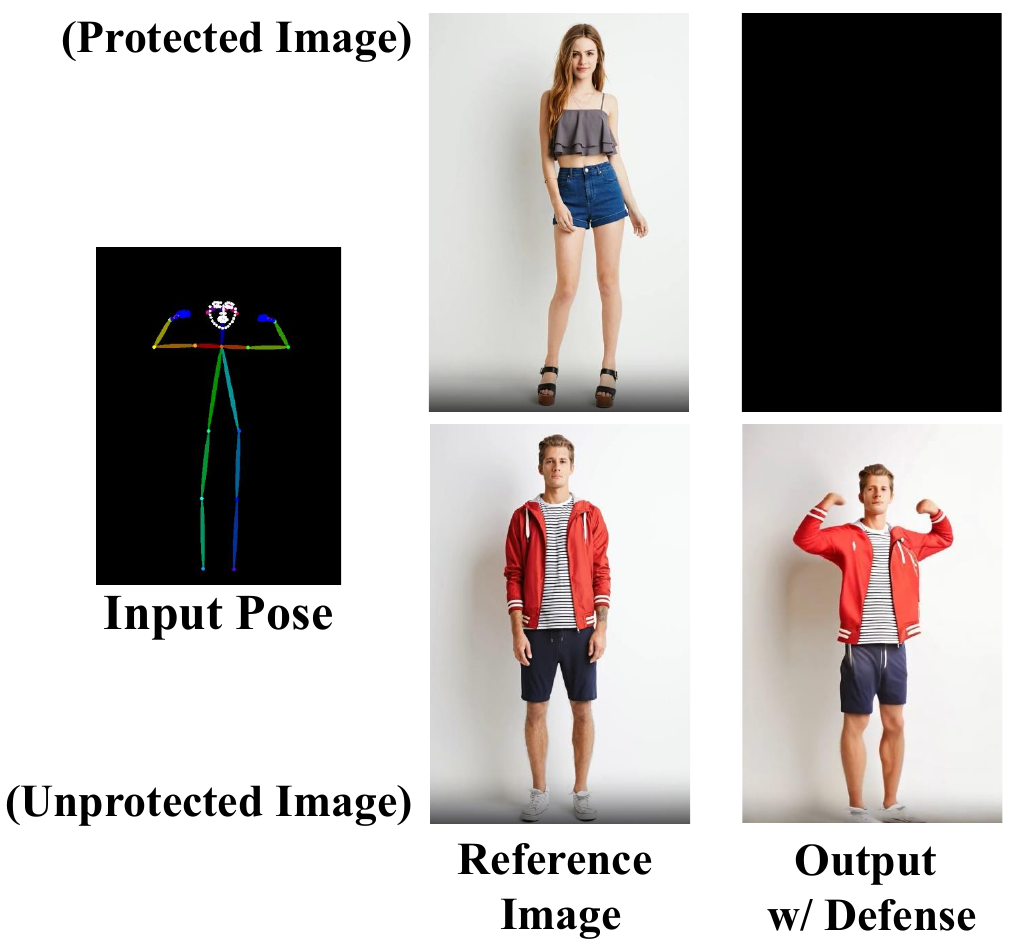}
   \caption{Visual examples of defense against reference image-conditioned generation}
   \label{fig:refimg}
\end{figure}

Beyond pose guidance, we extend PoseGuard to defend against reference image-conditioned video generation using the Animate Anyone model~\cite{hu2024animate}. As shown in Table~\ref{tab:refimg} and Figure~\ref{fig:refimg}, this strategy achieves substantially stronger suppression of unauthorized generation, with SSIM$^*$ dropping to 0.001. This improvement is attributed to the high concentration of identity-specific information in reference image representations, allowing the model to more effectively learn defensive perturbations. This demonstrates PoseGuard’s ability to prevent impersonation and unauthorized synthesis of specific individuals in reference image-driven scenarios.

\subsubsection{Defense against Facial Landmark-Guided Generation.}
\begin{figure}[t]
  \centering
   \includegraphics[width=0.8\linewidth]{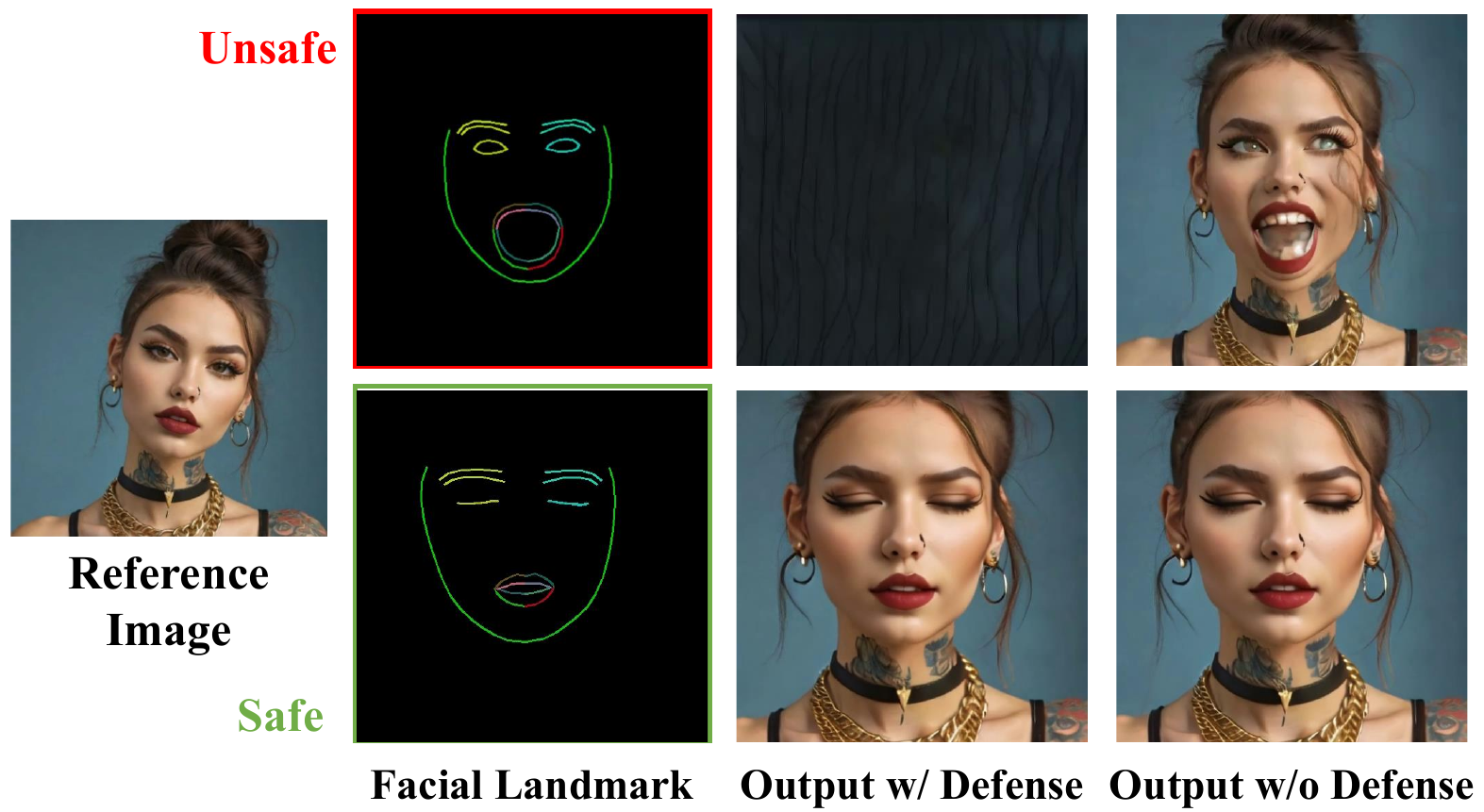}
   \caption{Defense visualization under facial landmark guidance. The top row denotes a harmful facial expression.}
   \label{fig:face}
\end{figure}

We further validate the generalizability of PoseGuard on AniPortrait~\cite{wei2024aniportrait}, a facial landmark-guided portrait video generation system. By fine-tuning the Denoising UNet with our defense objective, PoseGuard effectively suppresses unsafe facial landmarks, as illustrated in Figure~\ref{fig:face}. These results confirm PoseGuard’s flexibility across modalities, demonstrating its applicability to fine-grained facial manipulation tasks without compromising output quality for benign inputs.

\begin{figure}[t]
  \centering
   \includegraphics[width=0.85\linewidth]{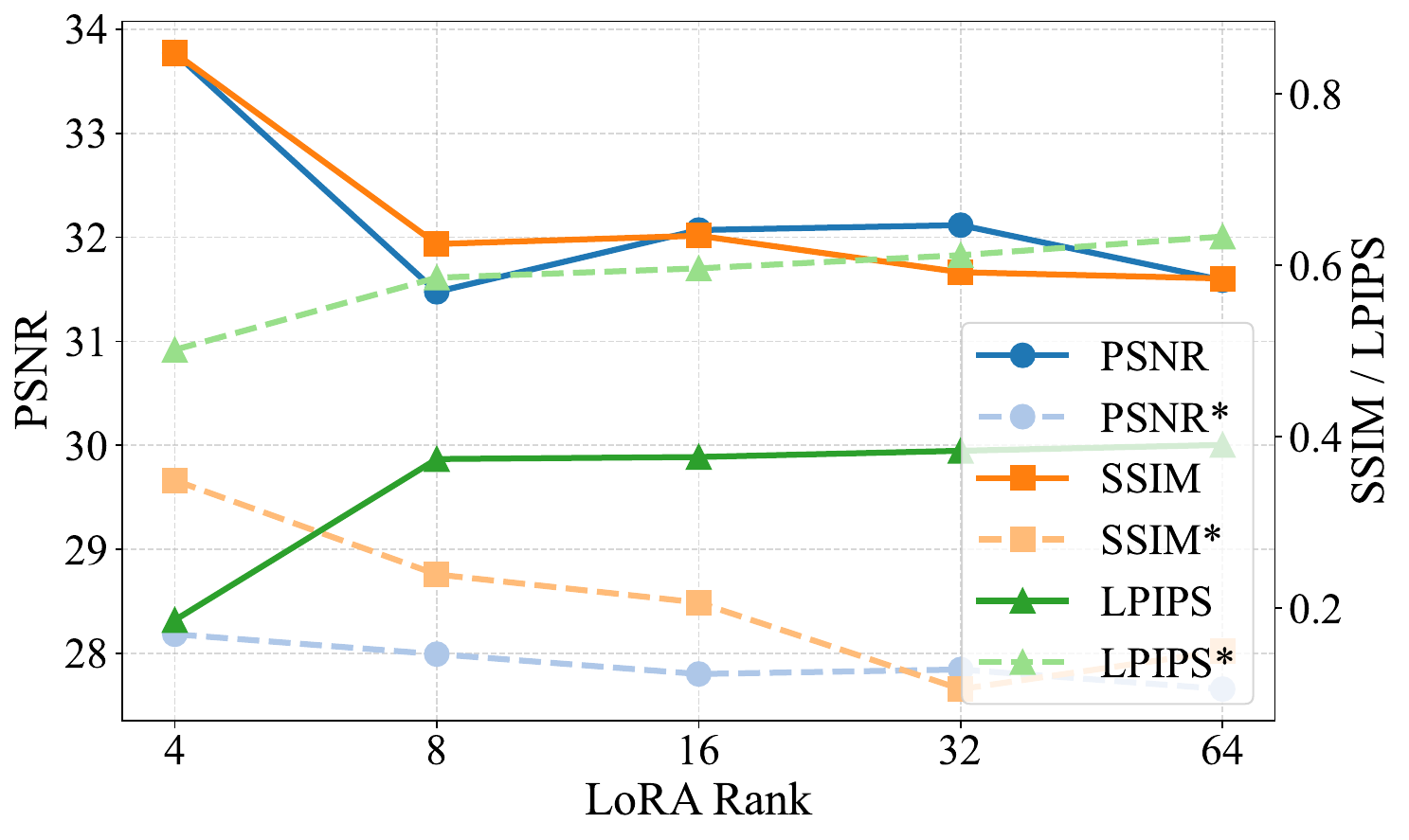}
   \caption{Impact of LoRA rank.}
   \label{fig:line_graph}
\end{figure}

\subsection{Ablation Studies}

\subsubsection{Effect of LoRA Rank.} As shown in Figure~\ref{fig:line_graph}, increasing LoRA rank from 4 to 64 slightly improves defense but significantly degrades benign generation quality (SSIM drops from 0.85 to 0.58). This validates that low-rank adaptation (rank = 4) is sufficient for effective suppression without compromising fidelity, while higher ranks cause overfitting and unnecessary quality loss.

\subsubsection{Safe Output Target Selection.}
\begin{figure}[t]
  \centering
   \includegraphics[width=\linewidth]{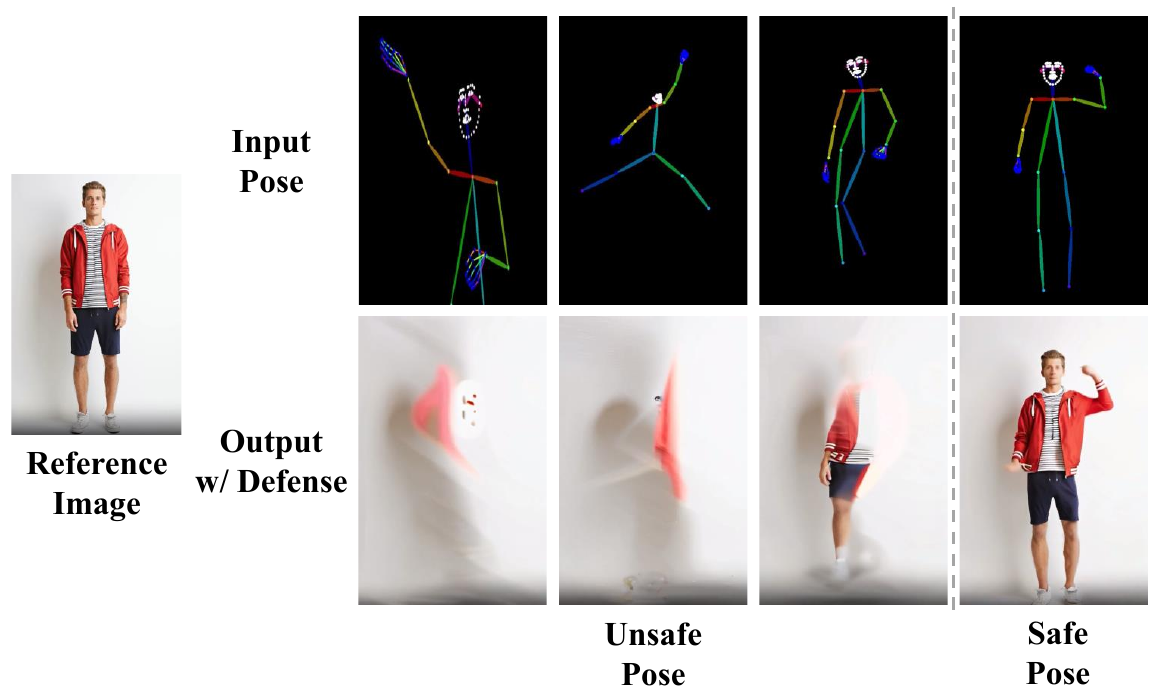}
   \caption{Visualization results of using blurred images as the safe output target.}
   \label{fig:blur}
\end{figure}

We compare black versus blurred images as safe targets. Blurred images better preserve benign visual quality while maintaining strong suppression, as shown in Figure~\ref{fig:blur}. This offers a flexible trade-off between defense strength and perceptual fidelity. See more details in the supplementary file.

\section{Conclusion}
We propose \textbf{PoseGuard}, a simple yet effective framework that embeds safety alignment into pose-guided generation models. PoseGuard selectively suppresses unsafe pose-conditioned outputs while maintaining high-quality generation for benign inputs, without additional inference overhead. Our approach offers a scalable and modular solution for safe content generation. Future work includes extending PoseGuard to broader multimodal generative scenarios, contributing to safer and more responsible AI systems.

\bibliography{reference}

\appendix

\clearpage

\section{More visualization Results}

\begin{figure*}
  \centering
   \includegraphics[width=\linewidth]{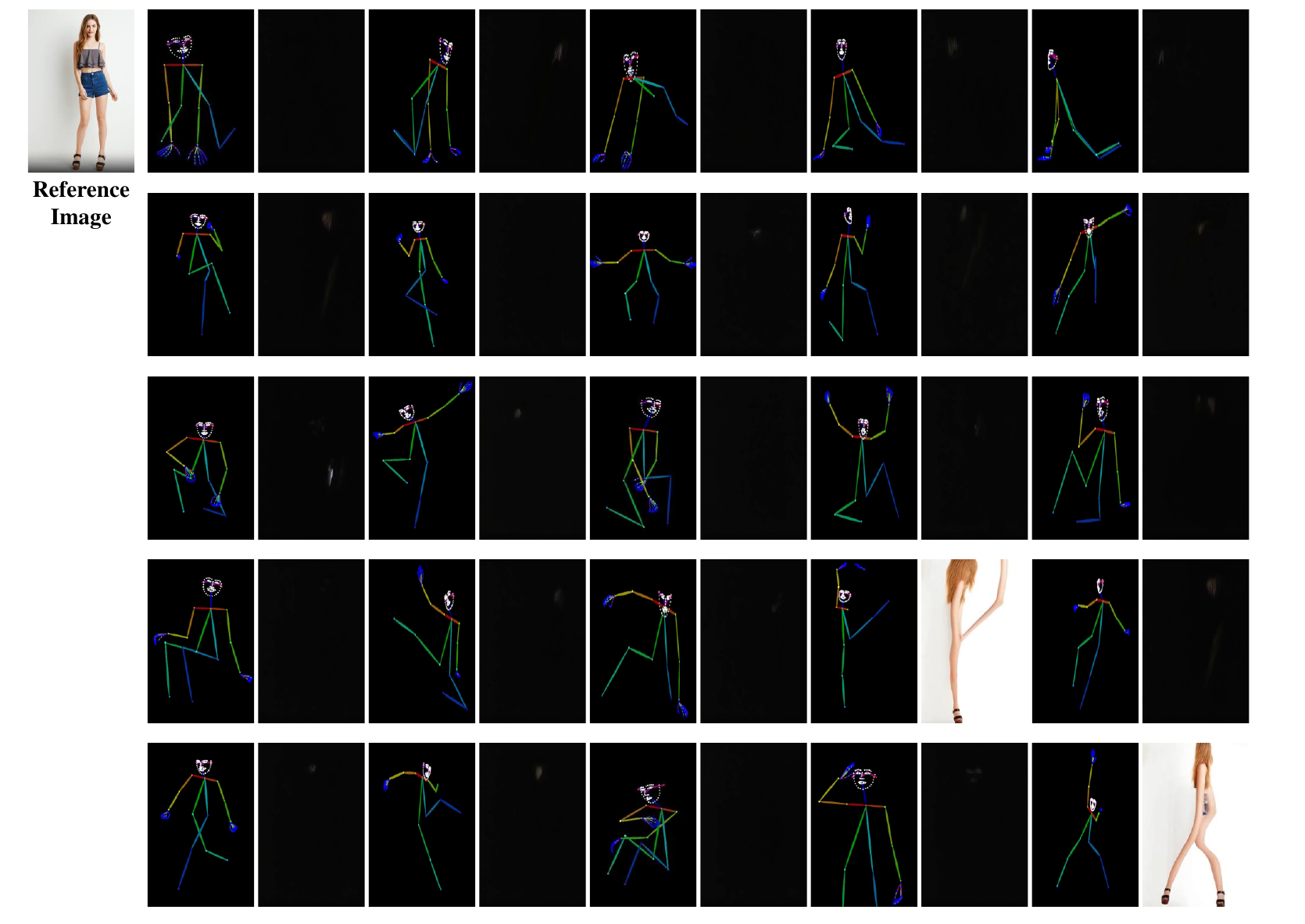}
   \caption{More Visualized Results of Full-parameter Fine-tuning.}
   \label{fig:more}
\end{figure*}

We provide more visualization results of \textbf{PoseGuard} based on full-parameter fine-tuning in Figure~\ref{fig:more}, covering various unsafe poses (e.g., discriminatory, NSFW) as well as any other poses \footnote{The poses used in this experiment are collected from an open-source platform Civitai.} that the defender intends to guard against. The visualizations showcase our method’s ability to effectively defend against a wide range of poses, causing the generation quality to severely degrade into either nearly black outputs or heavily distorted images. These results further demonstrate the strong pose capacity and defense effectiveness of \textbf{PoseGuard}.

\section{Results of Continuous Pose Sequences}

\begin{figure*}
  \centering
   \includegraphics[width=\linewidth]{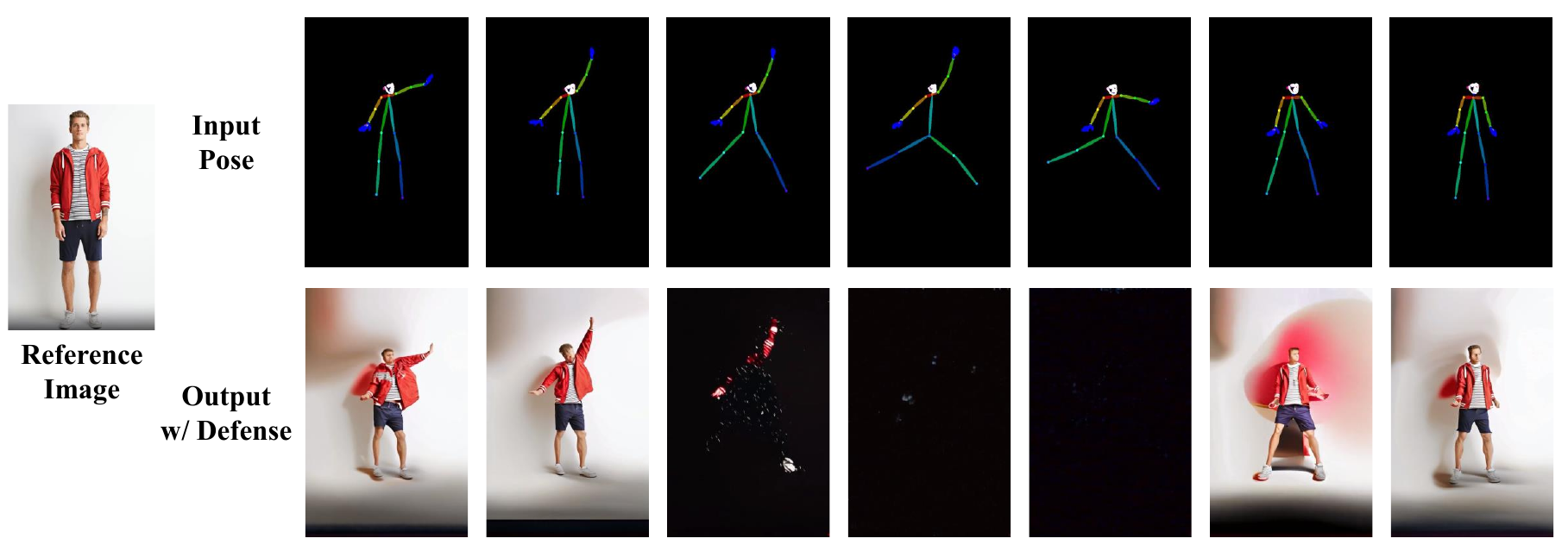}
   \caption{Results of continuous pose sequences.}
   \label{fig:seq}
\end{figure*}

Figure~\ref{fig:seq} shows more results of continuously varying pose sequences, demonstrating \textbf{PoseGuard}'s ability to handle transitional frames with unsafe poses embedded within otherwise normal pose sequences.

\section{Unsafe Pose Dataset}

\begin{figure*}
  \centering
   \includegraphics[width=\linewidth]{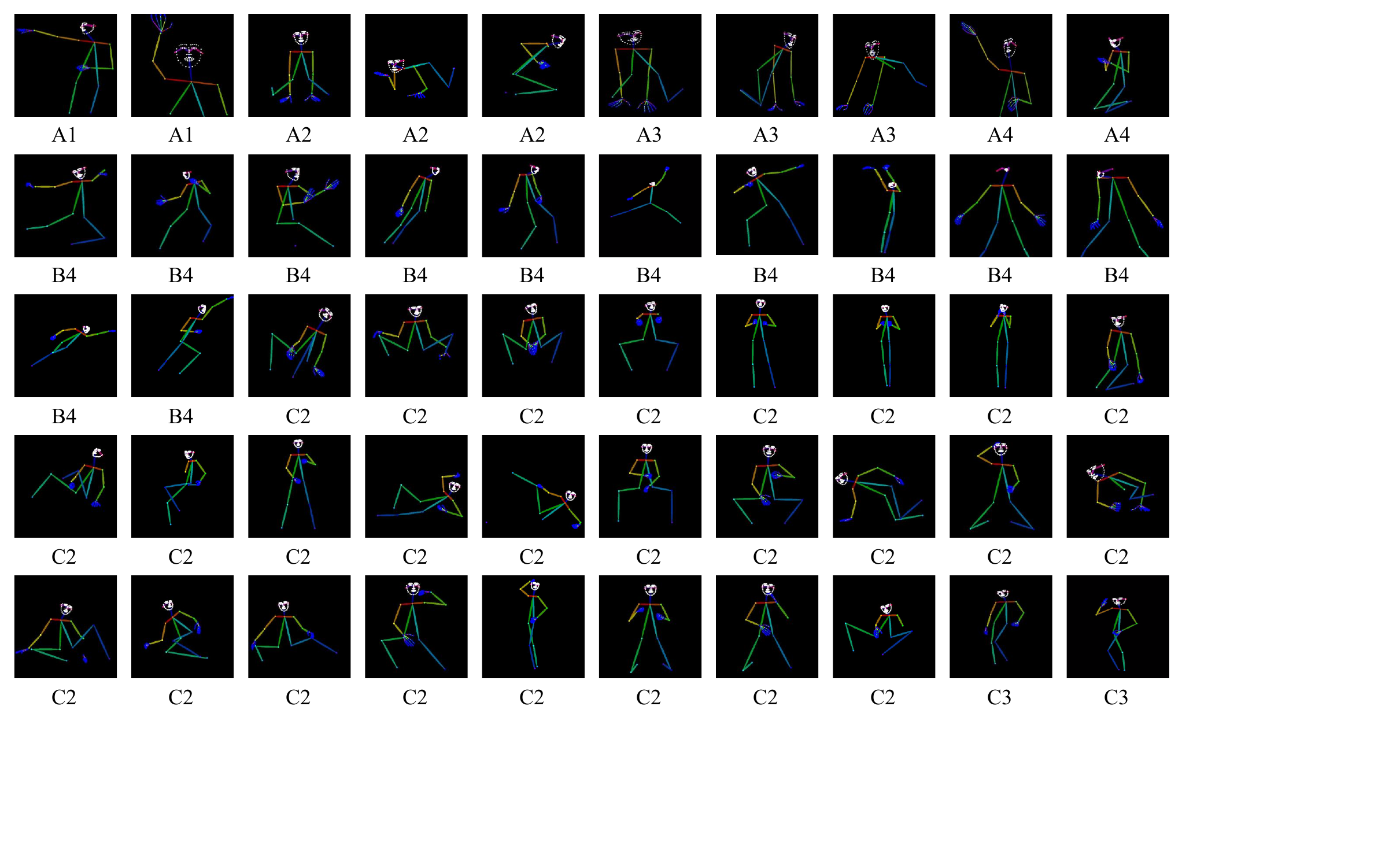}
   \caption{Examples of the 50 unsafe poses collected in our dataset. The poses are categorized into three categories: (A) Discriminatory pose, (B) Copyright-sensitive pose, and (C) NSFW pose. They are sourced from multiple online platforms, including (1) Wikipedia, (2) Render-State, (3) Civitai, and (4) Google Search.}
   \label{fig:poses}
\end{figure*}

Figure~\ref{fig:poses} shows examples of the 50 unsafe poses collected in our dataset. The full collection is available upon request for research purposes, subject to ethical review.

\section{Safe Output Target Selection}
We compare black versus blurred images as safe targets. Blurred images better preserve benign visual quality while maintaining strong suppression, as shown in Table~\ref{tab:blur}. This offers a flexible trade-off between defense strength and perceptual fidelity.

\begin{table*}
  \setlength{\tabcolsep}{1.8mm}
  \begin{tabular}{lcccccccccc}
    \toprule
    \multirow{2}{*}{Method} & \multicolumn{6}{c}{Generation Quality Metrics} & \multicolumn{4}{c}{Defense Metrics} \\
    \cmidrule(lr){2-7} \cmidrule(lr){8-11}
    & FID-VID $\downarrow$ & FVD $\downarrow$ & FID $\downarrow$ & SSIM $\uparrow$ & PSNR $\uparrow$ & LPIPS $\downarrow$ & SSIM* $\downarrow$ & PSNR* $\downarrow$ & LPIPS* $\uparrow$ & PSR $\uparrow$ \\
    \midrule
    LoRA-FT-4 & 7.842 & 231.10 & 31.863 & 0.88406 & 35.362 & 0.09740 & 0.79002 & 30.083 & 0.32498 & 1.000 \\
    LoRA-FT-8 & 12.800 & 278.77 & 43.752 & 0.87700 & 35.065 & 0.11023 & 0.79844 & 30.299 & 0.28938 & 0.958 \\
    LoRA-FT-16 & 19.690 & 327.40 & 60.563 & 0.86454 & 34.611 & 0.14182 & 0.77764 & 30.087 & 0.32567 & 0.980 \\
    LoRA-FT-32 & 18.974 & 292.13 & 54.128 & 0.86880 & 34.862 & 0.12998 & 0.76522 & 29.881 & 0.34330 & 0.980 \\
    \bottomrule
  \end{tabular}
  \caption{Results of using a blurred reference image as target safe output. LoRA-FT-$n$ denotes LoRA fine-tuning on $n$ unsafe poses.}
  \label{tab:blur}
\end{table*}

\section{Fine-Tuned Module Selection}
Table~\ref{tab:module} shows that fine-tuning both Denoising UNet and Reference UNet brings no notable gains over tuning only the Denoising UNet. We therefore adopt Denoising UNet fine-tuning to achieve strong defense with minimal parameter overhead.

\begin{table*}[t]
  \centering
  \begin{tabular}{lccccccccc}
    \toprule
    \multirow{2}{*}{Method} & \multicolumn{6}{c}{Generation Quality Metrics} & \multicolumn{3}{c}{Defense Metrics} \\
    \cmidrule(lr){2-7} \cmidrule(lr){8-10}
    & FID-VID $\downarrow$ & FVD $\downarrow$ & FID $\downarrow$ & SSIM $\uparrow$ & PSNR $\uparrow$ & LPIPS $\downarrow$ & SSIM* $\downarrow$ & PSNR* $\downarrow$ & LPIPS* $\uparrow$ \\
    \midrule
    FT-D & 18.750 & 281.71 & 28.957 & 0.87972 & 36.161 & 0.09418 & 0.06103 & 27.551 & 0.53305 \\
    FT-DR & 13.342 & 338.72 & 26.284 & 0.87760 & 35.511 & 0.08631 & 0.21349 & 27.719 & 0.55615 \\
    \bottomrule
  \end{tabular}
  \caption{Impact of fine-tuned modules on defense performance, under the setting of full-parameter fine-tuning with four poses while keeping all other parameters fixed. FT-D denotes fine-tuning Denoising UNet. FT-DR denotes fine-tuning Denoising UNet and Reference UNet.}
  \label{tab:module}
\end{table*}

\end{document}